# Deterministic influence of substrate-induced oxygen vacancy diffusion on $Bi_2WO_6$ thin film growth

Saikat Das[1,*], Tadakatsu Ohkubo[1], Shinya Kasai[1,2], and Yusuke Kozuka[1]

[1]Research Center for Magnetic and Spintronic Materials, National Institute for Materials Science (NIMS), 1-2-1 Sengen, Tsukuba 305-0047, Japan

[2]Japan Science and Technology Agency, PRESTO, Kawaguchi, Saitama 332-0012, Japan.

* Email: DAS.Saikat@nims.go.jp

**ABSTRACT**

In oxide epitaxy, the growth temperature and background oxygen partial pressure are considered as the most critical factors that control the phase stability of an oxide thin film. Here, we report an unusual case wherein diffusion of oxygen vacancies from the substrate overpowers the growth temperature and oxygen partial pressure to deterministically influence the phase stability of $Bi_2WO_6$ thin film grown by the pulsed laser deposition technique. We show that when grown on an oxygen-deficient $SrTiO_3$ substrate, the $Bi_2WO_6$ film exhibits a mixture of (001) and (100)/(010)-oriented domains alongside (001)-oriented impurity $WO_3$ phases. The (100)/(010)-oriented $Bi_2WO_6$ phases form a self-organized 3D nanopillar-structure, yielding a very rough film surface morphology. Oxygen annealing of the substrate or using a few monolayer-thick $SrRuO_3$ as the blocking layer for oxygen vacancy diffusion enables growing high-quality single-crystalline $Bi_2WO_6$ (001) thin film exhibiting an atomically smooth film surface with step-terrace structure. We propose that the large oxide-ion conductivity of $Bi_2WO_6$ facilitates diffusion of oxygen vacancies from the substrate during the film growth, accelerating the evaporation of volatile Bismuth (Bi), which hinders the epitaxial growth. Our work provides a general guideline for high-quality thin film growth of Aurivillius compounds and other oxide-ion conductors containing volatile elements.

# INTRODUCTION

Ever since the seminal work by Dijkkamp *et al.* demonstrating the synthesis of superconducting $YBa_2Cu_3O_{7-d}$ thin film,[1] oxide thin film growth by the pulsed laser deposition (PLD) technique has drawn widespread attention. This has led to the discovery of colossal magnetoresistance,[2] superconductivity in infinite-layer oxide thin film,[3] and ferroelectricity in an otherwise nonpolar oxide[4]. Furthermore, stacking dissimilar oxides in the form of heterostructure is a fertile playground for engineering the charge, lattice, and spin degrees of freedoms.[5] This enables stabilizing emergent functionalities like interface ferromagnetism,[6] metallic conductivity,[7,8] superconductivity,[9] and topological spin as well as polarization texture.[10,11] These are a few prominent examples constituting a highly active research field called "Oxide electronics", where PLD-grown oxide thin films play a central role.[12]

Stabilizing a desired epitaxial phase requires optimizing many parameters during the PLD growth. In general, temperature and background oxygen partial pressure are thermodynamically the most critical parameters, as is the case for all crystal growth techniques.[13,14] Characteristic of PLD, the laser wavelength, power density, and repetition rate of the pulsed laser affect the cation stoichiometry and microstructure of the film.[15] In addition to these growth parameters, equally important is the careful treatment of the substrate surface, which determines the initial growth sequence of oxide thin films and desired functionality.[8,16] In some cases, before thin film growth, the oxide substrates are *in-situ* annealed at high temperatures to obtain a singly terminated and atomically smooth surface.[17] Consequently, oxygen vacancies might be generated in the oxide substrate, which can impact the physical property of oxide thin film.[18] Controlling these defects is essential for realizing desired functionalities such as high-mobility two-dimensional electron gas.[19]

Here, we report an unusual impact of residual oxygen vacancies (OVs) in the substrate, which deterministically influences the phase stabilization of an oxide thin film. As a model system, we studied $Bi_2WO_6$—the primitive member of bismuth layered oxide Aurivillius compounds that exhibit the highest ferroelectric polarization within the Aurivillius family.[20] Besides, $Bi_2WO_6$ exhibits excellent visible-light photocatalytic property,[21] high oxide-ion conductivity,[22,23] edge conduction,[24] switchable photo-conductivity.[25] The in-plane nature of ferroelectricity/ferro-elasticity in $Bi_2WO_6$ also provides means to switch the chiral spin texture, such as magnetic vortex electrically.[26] Furthermore, the coupling between ferroelectricity and heavy elements like Bi and W could allow for ferroelectrically tunable Rasha spin-orbit coupling.[27] All these functionalities make $Bi_2WO_6$ an attractive system for spintronics applications. Previous studies focusing on the PLD growth of $Bi_2WO_6$ thin films on (001)-oriented $SrTiO_3$ reported strange growth behavior, whereby increasing growth rate promotes a mixed (001) and (113) oriented growth.[24,28] In this work, we show that residual OVs in $SrTiO_3$ promote polycrystalline thin film growth that consists of mixed (001) and (100)/(010)-oriented $Bi_2WO_6$ phases alongside the (001)-oriented impurity $WO_3$ phase. We argue that the diffusion of OVs from the substrate into the film during the growth leads to evaporation of volatile Bi species and triggers the phase decomposition. The epitaxial growth of high-quality (001)-oriented $Bi_2WO_6$ films is enabled either by refiling OVs through *in-situ* oxygen annealing of $SrTiO_3$ or blocking the diffusion of OVs by inserting a few monolayer-thick $SrRuO_3$.

**EXPERIMENTAL SECTION**

$Bi_2WO_6$ thin films were grown by PLD at 480–490 °C under an oxygen partial pressure of 120 mTorr. We employed the 4th harmonic ($\lambda = 266$ nm) excitation of an Nd:YAG laser operating at a repetition rate of 15 Hz and delivering about 6 mJ/pulse. Polycrystalline $Bi_{2.2}WO_6$ was used as the target. We employed single crystalline and chemically treated (001)-oriented $SrTiO_3$ substrates (Shinkosha Co., Ltd.). Prior to the growth, $SrTiO_3$ substrates were *in-situ* annealed at 950 °C under $10^{-6}$ Torr for about 30 minutes to obtain an atomically flat surface. Following the annealing step, $Bi_2WO_6$ films were grown using three different protocols. The first set of samples were grown by cooling down the substrate under $10^{-6}$ Torr to the growth temperature and subsequently increasing the background oxygen pressure to 120 mTorr (Sample A). For the second set of samples, the $SrTiO_3$ substrate was *in-situ* annealed for 45 minutes at 650 °C under oxygen partial 120 mTorr before cooling down to the growth temperature (Sample B). For the third set of samples, four monolayers of $SrRuO_3$ were grown on $SrTiO_3$ substrates at 650 °C under an oxygen partial pressure of 100 mTorr at a laser repetition rate of 2Hz with laser energy of about 20 mJ/pulse (Sample C). The temperatures outlined above were monitored using an optical pyrometer facing the substrate surface, and the emissivity was set to 0.8.

The crystal quality and the surface morphology of the films were characterized by X-ray diffraction (XRD) and atomic force microscope (AFM), respectively. The microstructure of the films was investigated using a transmission electron microscope (FEI Titan G2 80–200). For the scanning transmission electron microscopy (STEM) observations, an amorphous Au layer was sputtered on the film at room temperature. The amorphous Au prevents the damage during sample preparation using the focused ion beam and charging during STEM observation.

**RESULTS AND DISCUSSIONS**

Figure 1(a) displays the $2\theta$-$\omega$ XRD scan data of Sample A. The XRD pattern consists of very weak peaks of $Bi_2WO_6$ (004) and $Bi_2WO_6$ (006) alongside impurity peaks, which are centered at $2\theta = 23.02°$ and $47.3°$, respectively. These impurity peak positions are close to the cubic $WO_3$

(001) and (002) peaks.[29] The $c$-axis lattice parameter of $Bi_2WO_6$ estimated from the XRD data is 1.629 (3) nm, which is smaller than the bulk value ~ 1.643 nm.[30] Such contraction of $c$-axis lattice parameter could be accounted for in terms of the +1.4% in-plane tensile strain imposed by the $SrTiO_3$ substrate ($a$ = 0.3905 nm) on the $Bi_2WO_6$ film ($a$ = 0.5456 nm, $b$ = 0.5436 nm, pseudo-cubic lattice parameter = 0.3851 nm). Meanwhile, the lattice constant of the impurity phase is estimated to be 0.385(1) nm, comparable to the bulk lattice parameter of $WO_3$.[12] A rocking measurement performed around the $Bi_2WO_6$ (006) peak yields a full-width-at-half-maximum (FWHM) value ~ 0.1°, along with a weak secondary peak (Supplementary Figure S1(a)). Together with a very weak film peak, the rocking curve analysis, therefore, suggests poor crystallinity of the film. The AFM characterization reveals a homogeneous distribution of 3D particles (Fig 1(b)), yielding very rough surface morphology with a root-mean-square (RMS) roughness of about 4.4 nm.

In contrast to Sample A, a $2\theta$-$\omega$ XRD scan of the $Bi_2WO_6$ film grown on the oxygen annealed $SrTiO_3$ substrate (Sample B) reveals pronounced (001)-oriented $Bi_2WO_6$ peaks without any impurity phases (Fig. 1(c)). The rocking scan around the $Bi_2WO_6$ (006) peak shows an FWHM value of ~ 0.14° (Supplementary Figure S1(b)). The $c$-axis lattice parameter of this film is estimated to be 1.631 (3) nm, which is comparable to that of Sample A. The XRD data further shows clear thickness fringes covering a wide $2\theta$ range, which is indicative of a smooth film surface. The AFM measurement corroborates this conjecture and provides the evidence of an atomically smooth surface (RMS roughness ~ 0.38 nm) with a well-defined step-terrace structure (Fig. 1(d)). Overall, the combined XRD and AFM measurements clearly show that the oxygen annealing of the $SrTiO_3$ substrate enables the stabilization of the high-quality single-crystalline (001) phase.

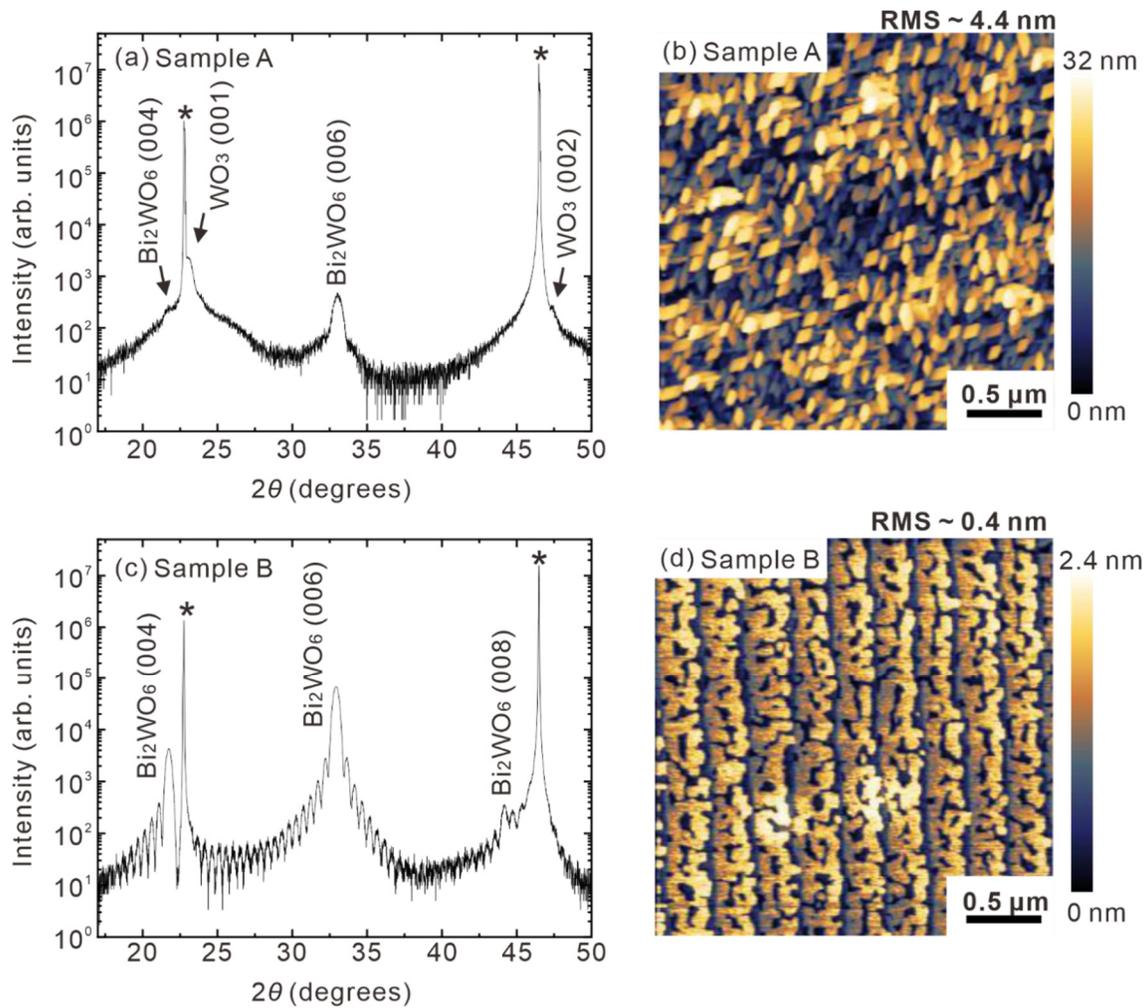

**Figure 1**. (a) XRD pattern and (b) AFM image of Sample A. The AFM image in (b) shows a rough surface with homogeneously distributed particles. The RMS roughness is about 4.4 nm. (c) XRD pattern and (d) AFM image of Sample B. AFM image in (d) shows an atomically smooth film surface, with a calculated RMS roughness of about 0.38 nm. In (a) and (b), SrTiO$_3$ peaks are marked by a star (*) symbol.

Next, to shed light on the difference between Sample A and Sample B on the microstructural level, in Fig. 2 we show cross-sectional high angle annular darkfield (HAADF)-STEM images taken along the SrTiO$_3$ [010] zone axis. The HAADF-STEM image of Sample A (in Fig. 2(a)) shows the layered $Bi_2WO_6$ phase (appearing in brighter contrast) coexists with a defective layer (appearing in black contrast). The inset figure, which magnifies the region enclosed by the red square, shows the layered $Bi_2WO_6$ phase is interfaced to a cubic phase through a $WO_4$ layer (marked by yellow spheres). Besides, the HAADF-STEM image also shows a vertical nano-pillar like structure (marked by a red arrow in Fig. 2(b)). The low magnification HAADF-STEM image shows these nano-pillar structures are homogeneously distributed over the film (Supplementary Fig. S2), and therefore, could be correlated to the 3D particles observed in the AFM image (Fig. 1(b)). A high magnification HAADF-STEM image further reveals the nano-pillar in Fig. 2(a) is surrounded by an amorphous layer (Fig. 2(b)). This amorphous phase is separated from the layered $Bi_2WO_6$ phase by grain boundaries (marked by white dashed-curves), which meet at the $Bi_2WO_6/SrTiO_3$ interface (marked by a triangle). Notably, the $Bi_2WO_6$ phase precedes by a diffused layer at the interface. Taking into account the expected $Bi_2WO_6$ growth sequence …-$WO_4$-$Bi_2O_2$-$WO_4$-$Bi_2O_2$/$TiO_2$-SrO-... on a $TiO_2$-terminated $SrTiO_3$ substrate, we, therefore, conclude that at the initial stage of deposition, the growth of the first $Bi_2O_2$ layer is unstable. In contrast to Sample A, the HAADF-STEM images of Sample B (Figs. 2(c)–2(d)) show a uniform film growth that is void of secondary phases and a reasonably abrupt $Bi_2WO_6/SrTiO_3$ interface.

To further understand the structure of Sample A in detail, we performed nanobeam electron diffraction (NBED) observation around a region, which consists of the layered $Bi_2WO_6$ phase and the vertically oriented nano-pillar structure. Figure 3(a) shows the corresponding HAADF-STEM image while Figs. 3(b)–(d) show the NBED pattern extracted from regions marked by red, green, and blue squares, respectively. The NBED pattern (Fig. 3(b)) from the area marked by the red square shows the characteristic diffraction pattern of a $c$-axis oriented $Bi_2WO_6$ domain. From the

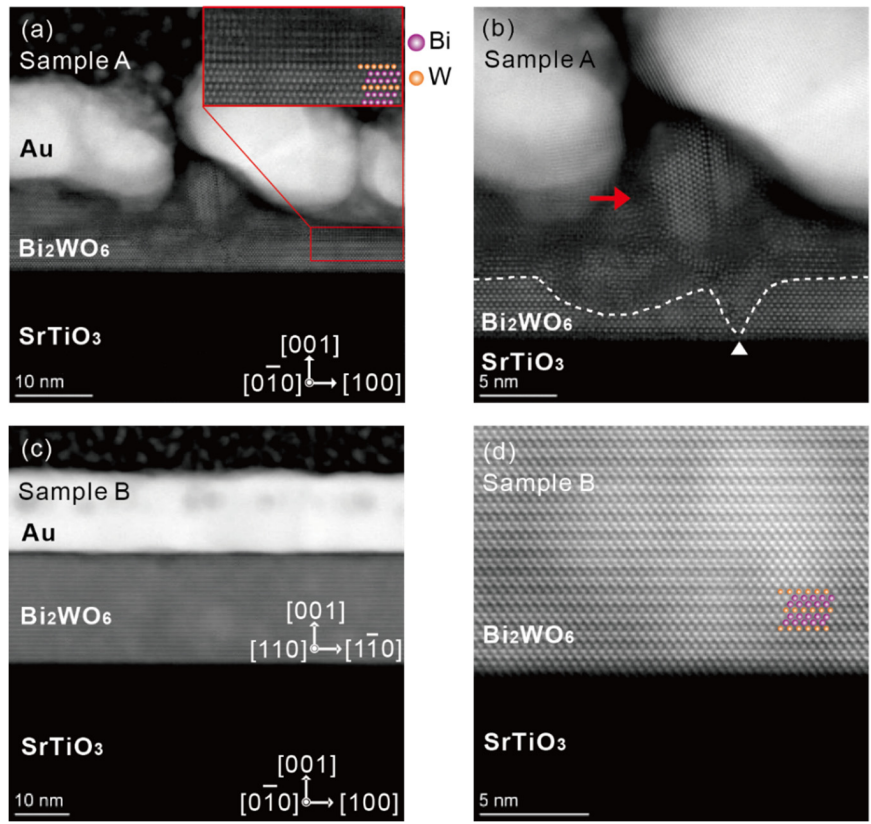

**Figure 2**. (a) HAADF-STEM image Sample A. The inset magnifies the boundary between the cubic and the layered phase. (b) High-magnification STEM image of sample A showing the grain boundaries and a diffused layer at the interface. The red arrow marks the vertical nano-pillar structure. (c) HAADF-STEM image of Sample B showing coherent film growth without visible defects and secondary phases. (d) High-resolution HAADF-STEM image showing a sharp $Bi_2WO_6$/$SrTiO_3$ interface in Sample B. To better illustrate the structure, in the inset of Fig. 2(a) and Fig. 2(d), we superimposed the model $Bi_2WO_6$ structure, where Bi and W atoms are shown as purple and orange spheres, respectively.

$(00\bar{6})$ and $(220)$ diffraction spots, we estimated the corresponding out-of-plane and in-plane spacings to be about 0.273 nm and 0.194 nm, respectively, which are reasonably comparable to the bulk values of 0.274 nm (= $c$/6) and 0.193 nm, respectively. The NBED pattern from the nano-pillar structure (Fig. 3(c)) appears to be rotated by 90° compared with that shown in Fig. 3 (b), suggesting the nano-pillars are either (100)- or (010)-oriented $Bi_2WO_6$ domains. In contrast, the NBED pattern from the region marked by the blue square exhibits a clear four-fold symmetry and very feeble ring-like feature surrounding the center. This implies that the region consists of mostly a cubic phase with a small contribution from non-crystalline, possibly an amorphous phase. The lattice spacing of the cubic phase is estimated to be around 0.387 nm, which is reasonably comparable to the lattice parameter of cubic $WO_3$,[12] and also consistent with the value estimated from the XRD measurement. Altogether, the STEM characterization unambiguously demonstrates the low crystallinity of Sample A as compared to Sample B. Besides providing the evidence of cubic $WO_3$ and non-crystalline phases, it also identifies the 3D particles seen in the AFM image of sample A as (100)- or (010)-oriented $Bi_2WO_6$ domains.

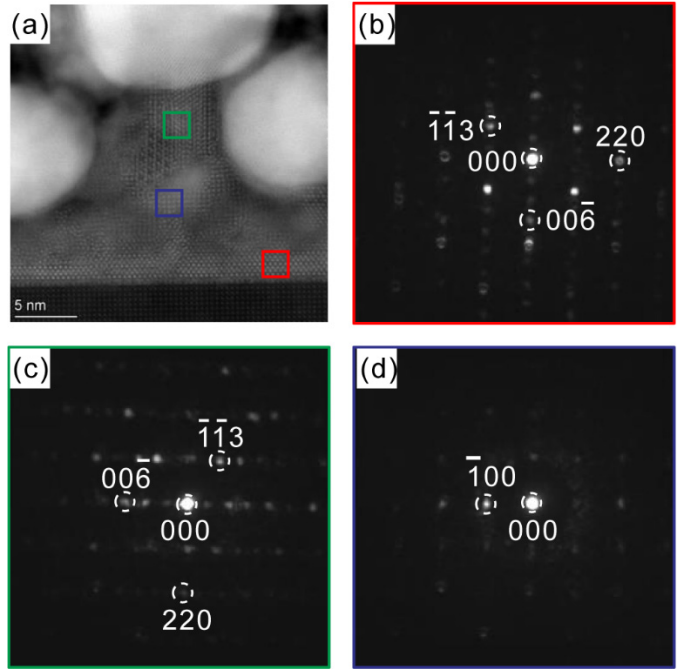

**Figure 3**. (a) High-magnification STEM image of Sample A around the nano-pillar structure. (b) NBED pattern from the red square region shows a (001)-oriented $Bi_2WO_6$ domain. (c) NBED pattern from the region marked by the green square showing (100)/(010)-oriented $Bi_2WO_6$ domain. (d) NBED pattern from the blue square region exhibits a mixture of cubic and amorphous like phases.

We now recall that both Sample A and Sample B were grown using identical PLD growth parameters, except that the $SrTiO_3$ substrate was *in-situ* annealed at 650°C under oxygen partial pressure of 120 mTorr before the growth of $Bi_2WO_6$ in the case of Sample B. This implies that the oxygen annealing step is crucial to stabilize the single-crystalline $Bi_2WO_6$ (001) phase and hints towards a possible role of OVs, which could be created during the annealing of $SrTiO_3$ substrates at 950°C under $10^{-6}$ Torr. To check this possibility, we treated two $SrTiO_3$ substrates, referred to as $SrTiO_3$ A and $SrTiO_3$ B hereafter, under the same conditions as used before the growth of Sample A and Sample B, respectively. This approach enables us to qualitatively probe the oxygen vacancies that might be prevailing in $SrTiO_3$ substrates before the $Bi_2WO_6$ growth. In supplementary Fig. S3, we show the pictures of $SrTiO_3$ A and $SrTiO_3$ B together with that of a pristine $SrTiO_3$ substrate. $SrTiO_3$ A appears to be darker than $SrTiO_3$ B, suggesting the presence of OVs.[31] From the room temperature Hall measurement, we estimated the carrier density in $SrTiO_3$ A to be around $2 \times 10^{20}$ cm$^{-3}$, which suggests that the approximate vacancy concentration in the substrate to be around $10^{20}$ cm$^{-3}$. In contrast, $SrTiO_3$ B showed insulating behavior. This implies that the oxygen annealing step effectively refills OVs formed during the high-temperature annealing in a vacuum. Meanwhile, OVs in large concentrations retain in the substrate that is exposed to oxygen at 500 °C.

Having confirmed that the $SrTiO_3$ substrate used for the growth of Sample A is nominally oxygen-deficient, we consider how it could trigger the collapse of the $Bi_2WO_6$ phase. We first note that Bi$_2$WO$_6$ is a good oxide-ion conductor like yttria-stabilized zirconia, which could support the facile migration of oxygen vacancies at the growth temperature (500 °C).[18] Secondly, we note that Bismuth has a very low melting point (~ 271 °C), and is extremely volatile, which makes Aurivillius oxide films are prone to Bismuth loss.[32] Meanwhile, the Tungsten oxide generally shows good immunity against oxygen vacancy-induced phase decomposition.[29] The XRD and STEM measurements (Figs. 1–3) demonstrate Sample A contains impurity $WO_3$ phases, which suggests the loss of Bi during the growth of this film. Based on these observations, we propose that due to the intrinsic high oxide-ion conductivity of $Bi_2WO_6$, OVs diffuse from the $SrTiO_3$ substrate into the film during the PLD growth. These OVs destabilize $Bi_2O_2$ layers, triggers the evaporation of Bi, and yields the impurity $WO_3$ phases. The observation of a diffused layer at the $Bi_2WO_6$/$SrTiO_3$ interface and grain boundaries connected to the interface in Sample A (Fig. 2(b)) could support this picture. These grain boundaries may act as the fast diffusion channels for OVs into the $Bi_2WO_6$ matrix.[33] The evaporation of Bi could also reduce the mobility of adatoms during the film growth. Accordingly, $Bi_2WO_6$ grows via Stranski-Krastanov growth mode (layer-by-layer followed by island growth), whereby 3D islands of (100)/(010)-oriented phase are formed on top of the layered (001)-phase, as seen in the STEM images (Figs. 2(a)-(b), and Fig. 3(a)). This picture is partly consistent with the previous reports, where a higher deposition rate leads to the formation of (113)-oriented 3D islands on top of the (001)-oriented layered phase.[34,25] The absence of high-density OVs in the $SrTiO_3$ substrate then naturally explains the high crystalline quality of Sample B.

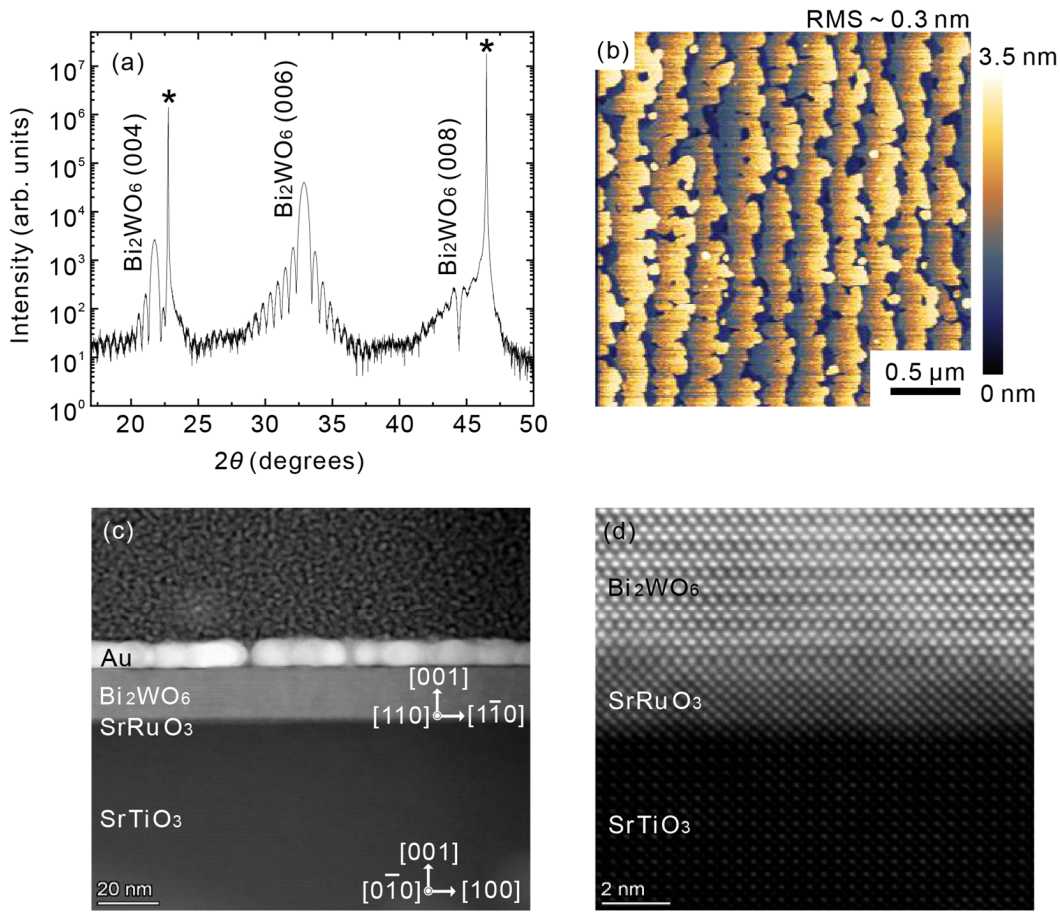

**Figure 4**. Structural characterization of Sample C. (a) XRD pattern of Sample C. SrTiO$_3$ peaks are marked by a star (*) symbol. (b) AFM image showing an atomically smooth film surface with a step-and-terrace structure. The RMS roughness amounts to about 0.28 nm. (c) Low-resolution HAADF STEM image showing the defect-free and coherently grown film. (d) High-resolution HAADF-STEM image showing atomically sharp Bi$_2$WO$_6$/SrRuO$_3$ heterointerface.

To further support our proposed mechanism, instead of the extended oxygen annealing step, we grew four monolayers of $SrRuO_3$ at 650 °C on the $SrTiO_3$ substrate before the deposition of $Bi_2WO_6$ film (Sample C). The diffusion coefficient of OVs in $SrRuO_3$ is among the lowest within the perovskite oxides,[35] and thereby can effectively block the migration of OVs from $SrTiO_3$ to $Bi_2WO_6$. Structural characterization of this sample by XRD, AFM, and STEM (Fig. 4 and Fig. S1(c)) unambiguously confirms that $Bi_2WO_6$ film grown on the $SrRuO_3$-buffered $SrTiO_3$ substrate is of high crystalline quality and exhibits layered structure with sharp $Bi_2WO_6$/$SrRuO_3$ interface (Figs. 4(b) and 4(c)).

**CONCLUSION**

In summary, we have shown that the phase stability of $Bi_2WO_6$ epitaxially grown on the $SrTiO_3$ substrate sensitively depends on the oxygen stoichiometry of the substrate. The presence of OVs in the substrate hinders the single crystalline (001)-oriented film growth and promotes the decomposition of film into impurity $WO_3$, amorphous, and polycrystalline $Bi_2WO_6$ phases. The polycrystalline $Bi_2WO_6$ phase consists of (001) and (100)/(010)-oriented domains, whereby the latter domains form a homogenously distributed vertically aligned nano-pillar structures. We propose that diffusion of oxygen vacancies from the substrate to $Bi_2WO_6$ and triggers excessive Bi loss and the concomitant collapse of the stochiometric phase. We further show that extended oxygen annealing of substrate or insertion of few monolayers of $SrRuO_3$ is effective in preventing the phase decomposition and enable high-quality thin film growth.

Our work uncovers a unique aspect of oxide thin film growth, wherein oxygen vacancies present in the substrate deterministically controls the phase stability of an oxide. It provides a guideline for stabilizing Aurivillius phases with desired stoichiometry and epitaxial orientation. Besides, our observation may be generalized for other oxides containing volatile species like bismuth (Bi)[36] and those with a large oxide-ion conductivity. Furthermore, it emphasizes the necessity to carefully control the oxygen stoichiometry of oxide substrates, which could be unintentionally overlooked during the optimization of oxide thin films.

ACKNOWLEDGMENT

This work was supported by Grants-in-Aid for Scientific Research (B) No. 19H02604. We would like to thank Mr. Jun Uzuhashi and Dr. Taisuke Sasaki for their assistance with the STEM measurements.

Abbreviations

| | |
|---|---|
| PLD | Pulsed laser deposition |
| OVs | Oxygen vacancies |
| XRD | X-ray diffraction |
| AFM | Atomic force microscopy |
| STEM | Scanning transmission electron microscopy |
| HAADF | High angle annular dark field |
| NBED | Nanobeam electron diffraction |

# Supporting Information

# Deterministic Influence of Substrate-Induced Oxygen Vacancy Diffusion on $Bi_2WO_6$ Thin Film Growth

Saikat Das[1,*], Tadakatsu Ohkubo[1], Shinya Kasai[1,2], and Yusuke Kozuka[1]

[1]Research Center for Magnetic and Spintronic Materials, National Institute for Materials Science (NIMS), 1-2-1 Sengen, Tsukuba 305-0047, Japan

[2]Japan Science and Technology Agency, PRESTO, Kawaguchi, Saitama 332-0012, Japan.

[*]Email: DAS.Saikat@nims.go.jp

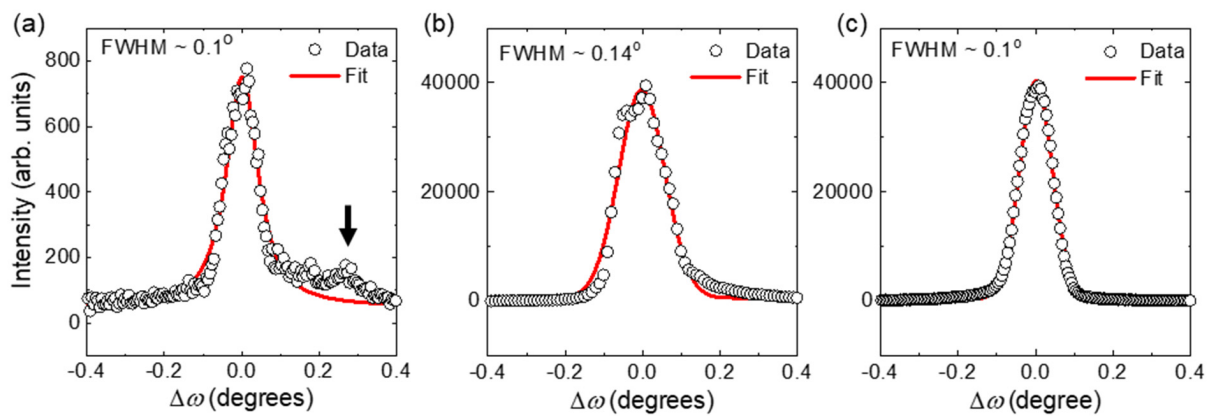

**Figure S1.** Rocking curves around $Bi_2WO_6$ (006) peak for (a) Sample A, (b) Sample B, and (c) Sample C. The solid line is a Pseudo-Voigt fit to the data. The vertical arrow in Fig. S1 (a) marks the secondary peak.

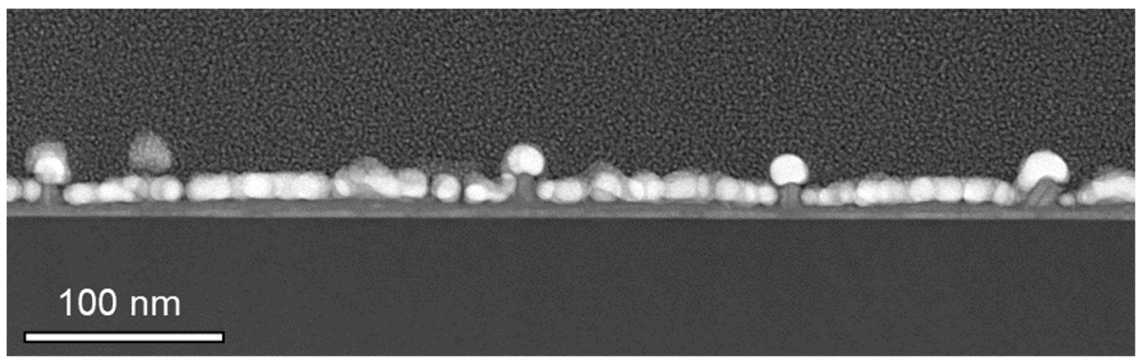

**Figure S2.** The low magnification HAADF-STEM image of Sample A, showing regularly arranged 3D nanopillar structures.

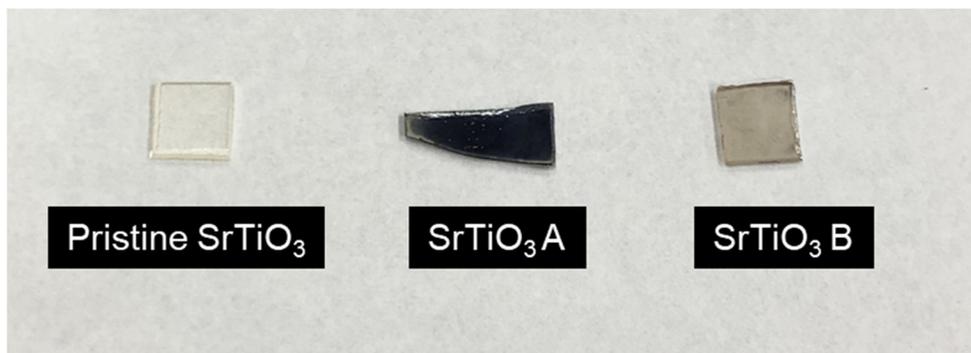

**Figure S3.** Optical image of SrTiO$_3$ A and SrTiO$_3$ B together with a pristine SrTiO$_3$ substrate.